\documentclass[11pt,logo]{preprint}
\usepackage{iftex}
\ifXeTeX
  \microtypesetup{tracking=false}
  \usepackage{newunicodechar}
  \newunicodechar{–}{\textendash}
\fi
\usepackage{tikz}
\usetikzlibrary{backgrounds,arrows.meta,positioning,calc}
\usepackage{xspace}
\usepackage{subcaption}
\usepackage{cite}
\definecolor{linkink}{HTML}{173B65}
\definecolor{nodeink}{HTML}{1F3859}
\definecolor{delink}{HTML}{96362E}
\definecolor{addink}{HTML}{1B6254}
\usepackage[colorlinks=true,allcolors=linkink,
  pdftitle={Absentia: Detecting Broken Access Control Vulnerabilities in Web Applications},
  pdfsubject={Research preprint}]{hyperref}
\usepackage{bookmark}
\usepackage{placeins}

\newcommand{\method}{\textsc{Absentia}\xspace}
\newcommand{\bench}{\textsc{BAC-Bench}\xspace}

\newcommand{\vanilla}{Plain LLM-Agent\xspace}
\newcommand{\result}[1]{#1}

\newcommand{\panelref}[2]{Figure~\hyperref[#1]{\ref*{#1}(#2)}}
\newcommand{\tinyquad}{\hspace{0.5em}}
\newcommand{\myparagraph}[1]{\par\noindent\textbf{#1}\enspace\ignorespaces}

\fancypagestyle{mainstyle}{%
  \fancyhf{}
  \fancyhead[C]{\footnotesize\textbf{\method: Detecting Broken Access Control Vulnerabilities in Web Applications}}
  \fancyfoot[C]{\footerfont\thepage}
  
  \renewcommand{\headrule}{\color{gray}\hrule height 0.5pt\vspace{1pt}}
}
\titleformat{\paragraph}[runin]{\normalfont\normalsize\bfseries}{}{0pt}{#1}
\titlespacing*{\paragraph}{0pt}{0.5\baselineskip}{0.5em}
\setlist[itemize]{topsep=4pt,itemsep=3pt,parsep=0pt}
\setlist[description]{style=nextline,leftmargin=0pt,labelindent=0pt,labelwidth=0pt,labelsep=0pt,
  topsep=4pt,itemsep=5pt,parsep=0pt}

\AtBeginEnvironment{thebibliography}{\raggedright}

\title{\method: Detecting Broken Access Control Vulnerabilities in Web Applications}
\author{%
  \begin{tabular}{@{}c@{}}
    Andr\'{e} Vicente Duarte\textsuperscript{1,2,*}\quad
    Aditya Oke\textsuperscript{1}\quad
    Rui Melo\textsuperscript{1,3}\\[4pt]
    Shubham Gandhi\textsuperscript{1}\quad
    Nachiket Kotalwar\textsuperscript{1}\quad
    Charmi Khandor\textsuperscript{1}\quad
    Danqing Wang\textsuperscript{1}\\[4pt]
    Arlindo L. Oliveira\textsuperscript{2}\quad
    Carolyn Ros\'{e}\textsuperscript{1}\quad
    Lei Li\textsuperscript{1}\\[8pt]
    \small\textsuperscript{1}Carnegie Mellon University\\
    \small\textsuperscript{2}INESC-ID, Instituto Superior T\'{e}cnico\\
    \small\textsuperscript{3}Faculdade de Engenharia do Porto
  \end{tabular}%
}
\correspondingauthor{Corresponding author:
  \href{mailto:andre.v.duarte@tecnico.ulisboa.pt}{\nolinkurl{andre.v.duarte@tecnico.ulisboa.pt}}}
\hypersetup{pdfauthor={Andr\'{e} Vicente Duarte; Aditya Oke; Rui Melo;
  Shubham Gandhi; Nachiket Kotalwar; Charmi Khandor; Danqing Wang;
  Arlindo L. Oliveira; Carolyn Ros\'{e}; Lei Li}}

\date{}

\begin{document}
\maketitle
\begin{abstract}
\noindent Broken access control, the failure of authorization, is one of the most prevalent web security risks. Unlike injection, a flow of untrusted input into a dangerous operation, authorization is a relation: who may act on what, not how data moves. Each application decides that relation for itself, so no rule written in advance carries to the next. An LLM agent can infer it from the code, but with no systematic way to cover the application and prioritize what to inspect, its search stays undirected and access-control flaws go undetected.
\par
We present \method, a security scaffolding that turns general LLM agents into systematic vulnerability detectors for the backend of web applications, run as an audit by the developers and security engineers who maintain the code. Under its direction, the agents build a graph that maps the application's routes to the code behind them. \method then works route by route, applying invariant falsification: it infers the properties the code is meant to satisfy, and where one is not enforced, reports the route for maintainer review.
\par
We also release \bench, a benchmark of 30 broken access control advisories across 25 repositories, 3 languages, and 9 frameworks, each published in 2025 or later, verified by a human auditor, and paired with its fixing commit, so credit requires flagging the vulnerable version and not the fixed one. \method recalls 19 of them, 17 under paired credit, and an LLM verifier confirms 51\% of its findings. CodeQL and Semgrep recall none, and an unstructured agent on the same model recalls 3. In the OWASP Benchmark injection categories, \method leads the dedicated analyzers in Python and trails only CodeQL and IRIS in Java. 

\end{abstract}

\section{Introduction}
\label{sec:introduction}

\begin{figure}[tbp]
  \centering
\begingroup
\tikzset{
  contrast/.style={x=1mm,y=-1mm},
  code/.style={anchor=base west,inner sep=0pt,
    font=\fontsize{7.8}{9}\selectfont\ttfamily,text=black!80},
  line/.style={anchor=base east,inner sep=0pt,
    font=\fontsize{7}{9}\selectfont,text=black!45},
  labeltext/.style={anchor=base west,inner sep=0pt,
    font=\fontsize{8}{10}\selectfont,text=nodeink},
  boxtext/.style={code,draw=nodeink!60,densely dotted,rounded corners=.4mm,
    inner xsep=.4mm,inner ysep=.65mm},
  annotation/.style={inner sep=0pt,
    font=\fontsize{7.5}{9}\selectfont\itshape,text=nodeink},
  pointer/.style={-{Stealth[length=.9mm,width=.65mm]},
    draw=nodeink!70,line width=.3pt},
}
\newcommand{\comparisonframe}[2]{%
  \fill[black!2,rounded corners=1mm] (0,0) rectangle (79.2,66);
  \fill[nodeink!8,rounded corners=1mm] (0,0) rectangle (79.2,10);
  \fill[nodeink!8] (0,8) rectangle (79.2,10);
  \draw[black!20,rounded corners=1mm] (0,0) rectangle (79.2,66);
  \node[anchor=west,text width=74.2mm,align=left,inner sep=0pt,
    font=\fontsize{8.8}{10.2}\selectfont\bfseries,text=nodeink]
    at (2.5,5) {#1};
  \node[labeltext,font=\fontsize{7.8}{9}\selectfont\ttfamily,text=black!60]
    at (7.6,13) {invoices.py};
  \node[anchor=base east,font=\fontsize{8}{10}\selectfont,text=nodeink]
    at (76.7,13) {#2};
  \draw[black!15] (5.6,15) -- (5.6,51);
  \draw[black!20,dashed] (2.5,56) -- (76.7,56);
}
\newcommand{\codeline}[3]{%
  \node[line] at (4.3,#2) {#1};
  \node[code] at (7.6,#2) {#3};
}
\newcommand{\diffline}[4]{%
  \node[line] at (4.3,#2) {#1};
  \node[code,text=#3] at (6.1,#2) {#4};
}
\begin{tikzpicture}[contrast]
  \comparisonframe{(a) Injection: unsafe source-to-sink flow}{CWE-89}
  \fill[delink!10] (5.8,31.7) rectangle (79,39.7);
  \fill[addink!11] (5.8,39.7) rectangle (79,47.7);
  \codeline{1}{18}{@app.get("/api/invoices/search")}
  \codeline{2}{22}{@login\_required}
  \codeline{3}{26}{def search\_invoices():}
  \node[line] at (4.3,30) {4};
  \node[code] (source-prefix) at (10.35,30) {q =};
  \node[boxtext] (source) at ([xshift=1mm]source-prefix.base east)
    {request.args["q"]};
  \node[annotation,anchor=west] (source-label)
    at ([xshift=4mm]source.east) {source};
  \draw[pointer] ([xshift=-.5mm]source-label.west)
    -- ([xshift=.3mm]source.east);
  \diffline{5}{34}{delink}{-}
  \node[code] (unsafe-prefix) at (10.35,34) {return};
  \node[anchor=base west,inner sep=0pt,font=\fontsize{7.8}{9}\selectfont,
    text=black!45] at ([xshift=2mm]unsafe-prefix.base east) {$\hookleftarrow$};
  \node[code] at (10.35,38) {db.execute(f"SELECT ... WHERE memo LIKE '\%\{q\}\%'")};
  \diffline{6}{42}{addink}{+}
  \node[code] (safe-prefix) at (10.35,42) {return};
  \node[anchor=base west,inner sep=0pt,font=\fontsize{7.8}{9}\selectfont,
    text=black!45] at ([xshift=2mm]safe-prefix.base east) {$\hookleftarrow$};
  \node[boxtext] (sink) at (10.35,46) {db.execute};
  \node[code] (sql-prefix) at ([xshift=.2mm]sink.base east)
    {("SELECT ... WHERE memo LIKE };
  \node[annotation,anchor=north] (sink-label)
    at ([yshift=-4mm]sink.south) {execution sink};
  \draw[pointer] ([yshift=.5mm]sink-label.north)
    -- ([yshift=-.3mm]sink.south);
  \node[boxtext] (parameter) at ([xshift=.7mm]sql-prefix.base east) {?};
  \node[code] at ([xshift=.2mm]parameter.base east) {", [q])};
  \node[annotation,anchor=north] (parameter-label)
    at ([yshift=-4mm]parameter.south) {parameterized};
  \draw[pointer] ([yshift=.5mm]parameter-label.north)
    -- ([yshift=-.3mm]parameter.south);
  \node[anchor=north west,inner sep=0pt,text width=73mm,
    font=\fontsize{8}{10}\selectfont\itshape,text=black!70] at (3,58)
    {Unsafe flow present: input reaches the query in a form that alters it.};
\end{tikzpicture}\hfill%
\begin{tikzpicture}[contrast]
  \comparisonframe{(b) Broken access control: authorization incorrectly enforced}{CWE-639}
  \fill[delink!10] (5.8,27.7) rectangle (79,35.7);
  \fill[addink!11] (5.8,35.7) rectangle (79,47.7);
  \codeline{1}{18}{@app.get("/api/invoices/<int:invoice\_id>")}
  \codeline{2}{22}{@login\_required}
  \codeline{3}{26}{def get\_invoice(invoice\_id):}
  \diffline{4}{30}{delink}{-}
  \node[code] at (10.35,30) {return Invoice.query.filter(};
  \diffline{5}{34}{delink}{-}
  \node[code] at (13.1,34) {Invoice.id == invoice\_id).first()};
  \diffline{6}{38}{addink}{+}
  \node[code] at (10.35,38) {return Invoice.query.filter(};
  \diffline{7}{42}{addink}{+}
  \node[code] (resourceprefix) at (13.1,42) {Invoice.id == };
  \node[boxtext] (resource) at ([xshift=.4mm]resourceprefix.base east) {invoice\_id};
  \node[code] at ([xshift=.4mm]resource.base east) {,};
  \node[annotation,anchor=west,font=\fontsize{7}{9}\selectfont\itshape]
    (resource-label) at ([xshift=5mm]resource.east)
    {requested resource};
  \draw[pointer] ([xshift=-.5mm]resource-label.west)
    -- ([xshift=.3mm]resource.east);
  \diffline{8}{46}{addink}{+}
  \node[boxtext] (ownership) at (13.1,46) {Invoice.account\_id == current\_user.id};
  \node[code] at ([xshift=.4mm]ownership.base east) {).first()};
  \node[annotation,anchor=north] (ownership-label)
    at ([yshift=-4mm]ownership.south) {ownership check};
  \draw[pointer] ([yshift=.5mm]ownership-label.north)
    -- ([yshift=-.3mm]ownership.south);
  \node[anchor=north west,inner sep=0pt,text width=73mm,
    font=\fontsize{8}{10}\selectfont\itshape,text=black!70] at (3,58)
    {Required check absent: ownership of the invoice is never established.};
\end{tikzpicture}
\endgroup

  \caption{
    Two vulnerabilities behind the same \texttt{@login\_required}
    authentication check.
    (a) SQL injection (CWE-89): a request parameter reaches the query
    unsafely.
    (b) Broken access control (CWE-639): the handler returns an invoice
    without verifying it belongs to the caller.
    The first is a flow a conventional taint rule can trace; the second
    depends on an application-specific ownership relation that such a rule
    does not encode. \method detects both.
  }
  \label{fig:crown-figure}
\end{figure}
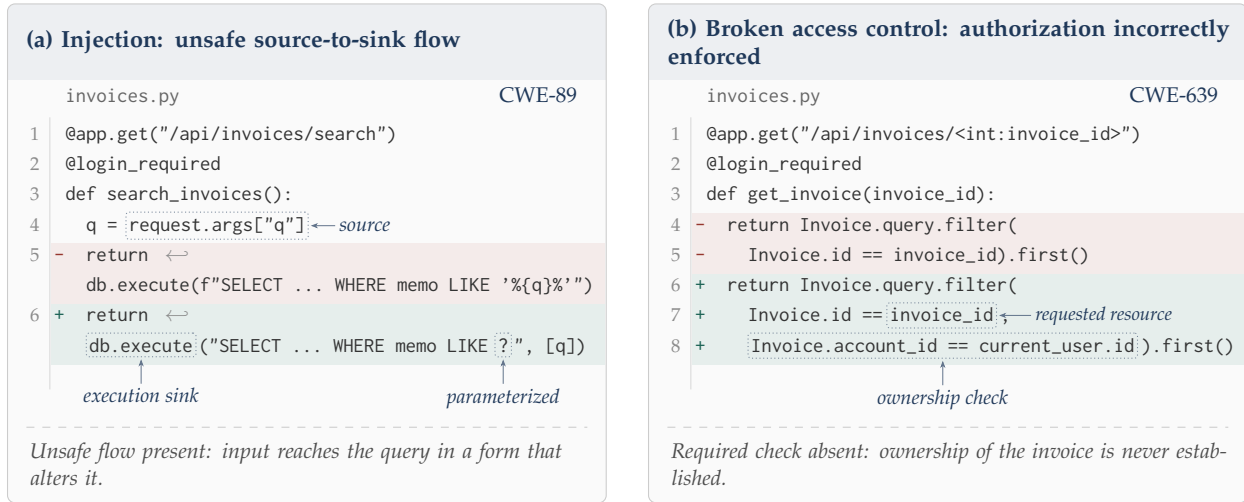

OpenReview is a platform on which many machine-learning conferences run peer review. In November 2025, during ICLR 2026's review cycle, a broken access control flaw in its API exposed the identities of authors, reviewers, and area chairs for more than 10,000 submissions~\cite{iclr2026incident}. Parties who obtained the data then harassed, intimidated, and offered bribes to reviewers, and the organizers could no longer establish which reviews had already been influenced. Scores were restored to their pre-breach state and the area chair of every submission was reassigned, but the anonymity that the process depends on could not be recovered.

A comparison with SQL injection shows what makes broken access control hard to detect in source code (Figure~\ref{fig:crown-figure}). In an injection the defect is a flow, in which a request parameter reaches the database query in a form that lets it alter the statement. The property violated is the same in every application, so static application security testing (SAST) tools trace these flows with rules written once for a language. Broken access control has no flow to trace, because the defect is a relation between a principal and a resource: who has authorization to act on what. A route is vulnerable when that relation is not enforced or is enforced incorrectly.
\par

SAST analyzers such as CodeQL and Semgrep~\cite{codeql,semgrep} carry authorization rules, but these test the presence of a conventional framework-specific guard, which is a syntactic fact that a pattern can match. Whether it is the right guard for the object that returns the route depends on the intended policy of the application under analysis, so no generalizable set of patterns can be written, and the class remains largely beyond the reach of the analyzes that make injection detectable~\cite{mocguard2025}. Detecting a missing or wrong check requires knowing which one the operation needs, and today that knowledge comes from one of two sources. The first is dynamic testing, which sends requests to the running application on different accounts and compares the responses~\cite{owasp-wstg-42,zap,bacscan,bacfuzz}. The second is manual review, which recovers the knowledge directly but requires an engineer to read the source route by route. Checking authorization from source is therefore a reading task. This paper asks whether that reading can be automated: whether the property each route is meant to guarantee can be recovered from the source and the code held to it, so that a reviewer verifies findings instead of performing the analysis.
\par
Large language models (LLMs) are currently the closest automated approximation to an engineer reading code. A model can read a route's handler together with the data model behind it and the conventions of the framework it is built on, and infer from them what that route is meant to enforce. However, left to read without direction, a model judges security properties unreliably ~\cite{secholmes}. \method is a scaffolding that supplies the direction in two steps: first, it maps the application's request surface into a graph of routes and the guards in front of them, and then audits that surface one route at a time by \emph{invariant falsification}, inferring the properties each route is meant to guarantee and checking the code against them. A route where an invariant fails becomes a finding for a security reviewer to verify.

\par
To validate \method, we assemble a benchmark of real authorization vulnerabilities: 30 disclosed advisories spanning 25 open-source applications and 9 frameworks, in Python (18), TypeScript (9), and JavaScript (3). Each advisory was published in 2025 or later, which limits what an evaluated model could have seen in training (\S\ref{sec:discussion-limitations}), and every instance was verified by a human auditor. Each instance carries a vulnerable and a fixed commit, so recall is scored together with a patched-commit differential that credits a detection only when the defect is reported on the vulnerable version and not on the fixed one. To assess whether \method is specific to authorization, we also run it on the injection categories of the OWASP Benchmark in Python and Java ~\cite{owasp-benchmark-python,owasp-benchmark-java}.

\par

The experiments yield three results. Broken access control is detectable from source: \method recalls 19 of the 30 vulnerabilities, 17 of them under the patched-commit differential, at 51\% verifier-adjudicated precision, while CodeQL and Semgrep recall none, an outcome consistent with a limit of expressiveness rather than of reach. What separates \method from a model left to itself is direction rather than capability: an unstructured agent on the same model, run on the same repositories as a control, recalls 3 of the 30. The same configuration reaches beyond that class: on the injection categories of the OWASP Benchmark it leads every tool we run in Python and trails only CodeQL and IRIS in Java. Together, they attribute the detection to the procedure rather than to the model: an analysis that recovers, route by route, the property the application intended to enforce, and treats it as a claim to falsify. The scaffolding converts a model's capacity to read code into that analysis.

Our main contributions are as follows:
\begin{itemize}
  \item \textbf{We propose \method, a scaffolding that directs general LLM agents to analyze source code systematically}, built on request-surface mapping, per-route invariant falsification, and the composition of findings across routes. Because what a route must guarantee is recovered from the application under audit rather than written in advance, \method runs unchanged on frameworks and languages it was not built for, and it reaches broken access control, a class that dataflow analyses miss in most cases.

  \item \textbf{We construct and release \bench, a broken access control benchmark}: 30 disclosed advisories across 25 repositories, 3 languages, and 9 frameworks, each published in 2025 or later, verified by a human auditor against its advisory and fix commit, scored by recall and a patched-commit differential.

  \item \textbf{Our experiments show that scaffolding turns a general agent into a detector}: \method recalls 19 of the 30 vulnerabilities at 51\% verifier-adjudicated precision, where CodeQL and Semgrep recall none and an unstructured agent on the same model recalls 3. When the components are separated, the per-route decomposition accounts for the recall, and invariant falsification cuts report volume by a third at no cost to it. We also report costs and how much of previous computations can be avoided when a repository is re-analyzed after a commit.
\end{itemize}

\section{Background}
\label{sec:background}
\label{sec:two-families}

\par
Web-application vulnerabilities fall into two families~\cite{logic-vulns-usenix-2010,static-analysis-usenix}: \emph{common} flaws, detectable without any model of what the application is meant to do, and \emph{application-specific} flaws, invisible without one. Injection is the standard example of a common flaw. SQL injection, cross-site scripting, and command injection share a common abstraction: untrusted input reaches a dangerous operation without being made safe first \cite{sql-injection-livshits}. That definition is written once for a language and its libraries, by naming which functions read input from a request and which operations must not receive that input unchecked. It then applies unchanged to every application written in that language. The defect is a property of a dataflow between those two sets of functions, and this is the \emph{taint signature} that an analysis can trace directly.
\par
Broken access control (BAC, OWASP A01:2025~\cite{owasp-web-top10-25}) is the failure of \emph{authorization}: the policy that decides which principals may act on which resources. We write that policy as a relation, $\mathit{permitted}(u,r)$, asking whether principal $u$ may act on resource $r$ \cite{mocguard2025}. Unlike the source and sink specifications used by conventional taint analyses, the relevant authorization relation is usually application-specific. A taint analysis can determine whether request-derived data reaches a sensitive operation, but it does not by itself encode whether that operation is permitted for this principal and resource. Detecting the defect therefore requires either an application-specific policy specification or a procedure that recovers a candidate policy from the application under analysis.
\par
Authorization failures take two forms: a required check is absent, or a check is present but enforces the wrong principal, resource, role, or scope. The presence of a guard is therefore insufficient; correctness depends on whether it implements the application's intended policy.

\section{Motivation}
\label{sec:overview}

Broken access control is hard to detect in source code because the property a route must guarantee is particular to the application under audit, while an injection defect is defined once for a language and its libraries (\S\ref{sec:background}). We first state the setting in which that property has to be recovered, and the failures we claim to detect. A short example then shows that difference inside one application, and three obstacles follow from it. Those obstacles imply three requirements, and \S\ref{sec:design} presents the scaffolding built to meet them.

\subsection{Setting and Scope}
\label{sec:scope}

\myparagraph{Setting.} \method is run by an application's own developers and security engineers, as an audit of code they maintain. White-box access is therefore the natural setting for it. \method reads the application's source code, with no running instance, no credentials, and no live endpoint. It reports the access-control defects it finds so that they can be fixed. A defect only has meaning relative to a principal, so we fix one reference point. A violation is what an authenticated but low-privilege principal, for instance an ordinary user, can do beyond what that principal is permitted: a request for which $\mathit{permitted}(u,r)$ fails. Finally, \method detects rather than exploits: it reports what a request would have to do to violate the policy, which is what a maintainer needs in order to confirm the defect and fix it.

\myparagraph{Scope.} \tinyquad We take broken access control in the sense of \S\ref{sec:two-families}. A route falls in scope when the application would accept a request that reaches data or functionality the requesting principal is not entitled to, whether no check stands in the way or one does and enforces the wrong relation. The advisories behind our instances name this failure in two main ways. One is CWE-639, where a request supplies the identifier of an object another principal owns and the application does not verify that the caller is entitled to it, commonly called an insecure direct object reference (IDOR). The other is CWE-862, missing authorization, where an operation performs no entitlement check at all. \S\ref{sec:benchmark} reports how the instances fall across these labels.
\par
Run unchanged, \method also reaches the injection defects of \S\ref{sec:two-families}, which \S\ref{sec:evaluation} reports as evidence that it is not specific to authorization. Two classes commonly grouped with access control fall outside this scope because neither is a failure of the relation above. The first is defeating authentication itself, which includes credential theft, session or token forgery, and abuse of login and password reset. The second is compromise of the infrastructure the application runs on.

\subsection{Two Vulnerabilities Behind One Guard}

Figure ~\ref{fig:crown-figure} shows two vulnerabilities from the same application, reduced to their essential shape. Both routes are declared in the same file, and both carry the same authentication check, \texttt{@login\_required}, on line 2. In \panelref{fig:crown-figure}{a} the caller controls the query parameter \texttt{q}, which line 5 inserts into the SQL string handed to \texttt{db.execute}, so the caller can change the query itself, not only the text it searches for. \panelref{fig:crown-figure}{b} takes an invoice identifier from the request path and returns the invoice carrying that identifier (line 4). Invoices belong to accounts in this application. The \texttt{@login\_required} check passes for any logged-in caller, and the query selects the invoice by identifier alone, so nothing confirms that the invoice belongs to the caller's account, and any authenticated user can read any invoice by supplying its identifier. The first defect is the kind most static analyzers report today and the second is not. Three obstacles separate them.

\subsection{Obstacles and Requirements}

\myparagraph{O1: The property to be checked is not the same in every application.} An injection rule is written before the application exists: user input must reach a query only in a form that cannot alter it, in \panelref{fig:crown-figure}{a} as in any codebase. Nothing plays that role in \panelref{fig:crown-figure}{b}, whose handler would be correct, unchanged, in an application where invoices are shared documents. What makes it a defect is a property of this application alone: an invoice has an owning account, and only that account may read it. No rule carries that property in from outside, so it must be established for this route, from the code of this application, before that code can be checked against it. This is why the static analyzers we run report none of the 30 vulnerabilities in \bench (\S\ref{sec:evaluation}).

\myparagraph{O2: A missing check is a claim about the whole application, not about a line.} A rule can find the query on line 4 of \panelref{fig:crown-figure}{b}, since a filter that comes from the request is a pattern like any other. Whether it is a defect depends on something the line does not contain: it is wrong only if nothing anywhere else has already kept the caller to their own invoices. Unable to settle that, a rule can only report every read by identifier. What an analysis needs first is therefore the application's request surface, wide enough that the absence can be established across all of it. Applications also differ in how they declare routes, attach guards, and reach their data, and those in our benchmark span nine frameworks and three languages (\S\ref{sec:benchmark}), so a representation built around the conventions of one framework reaches only the applications that follow them.

\myparagraph{O3: A model reading the same code has the capacity but no procedure.} A language model can read a route's handler and the data model behind it and work out what that route is meant to allow. What it does not have is a way to work through an application. Two things are missing. The first is coverage: the applications in our benchmark expose a median of 220 routes and as many as 999 (\S\ref{sec:benchmark}), and an agent handed a repository and asked to look for access-control flaws reads whichever parts draw its attention, leaving no record of what it did not read. The second is a standard against which to judge: on a route it does read, whether the code looks wrong has no fixed answer, and the models decide from the surface features of the code rather than from what the route was meant to allow~\cite{secholmes}. In our evaluation, an unstructured agent using the same model detects 3 of 30 advisories, compared to 19 for \method (\S\ref{sec:evaluation}). The result motivates testing whether explicit coverage, rather than the underlying model alone, accounts for the difference.

\begin{figure}[tbp]
    \centering
    \includegraphics[width=\linewidth]{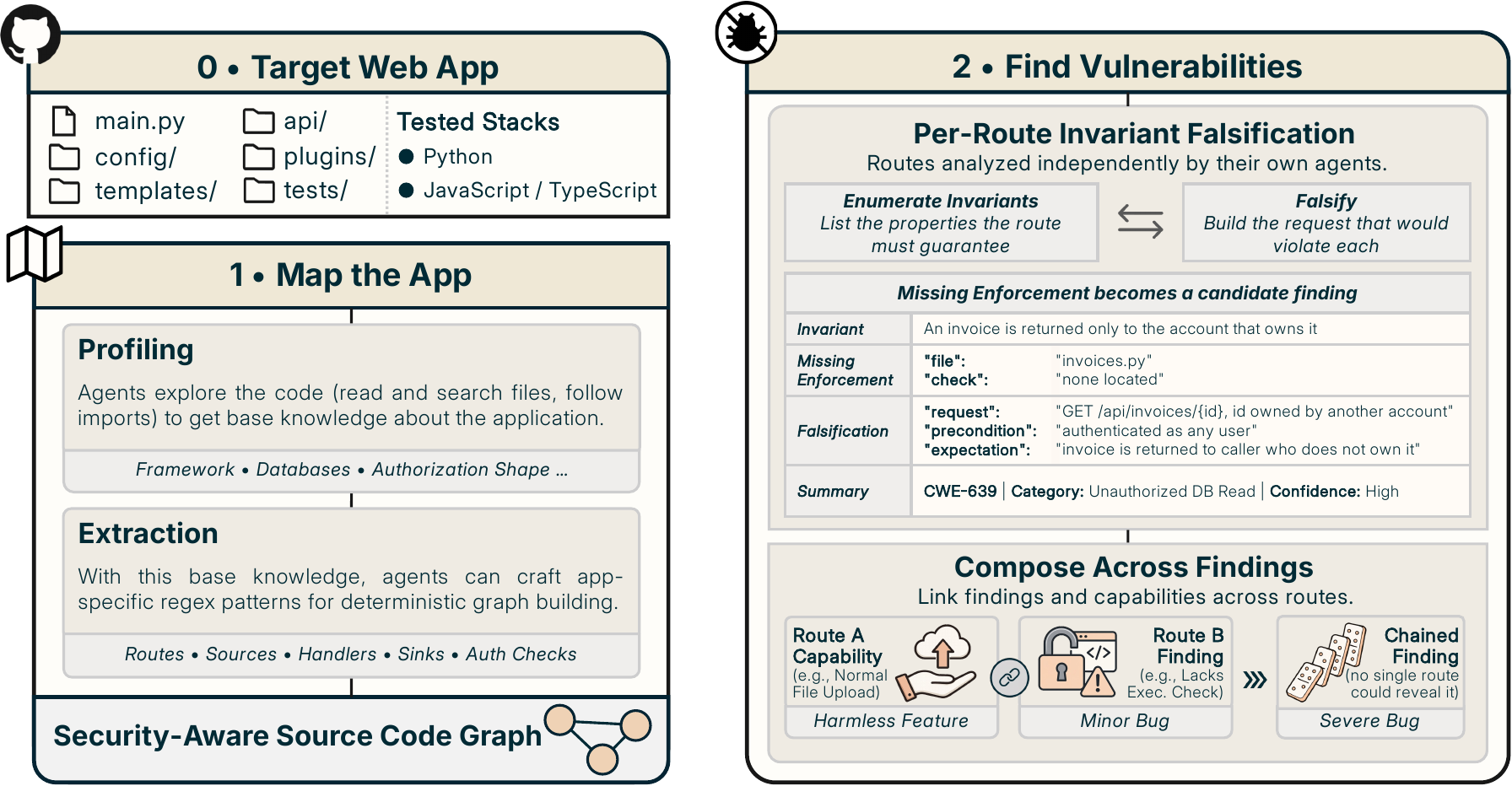}
    \caption{The stages \method runs over a target application: agents map the application into a graph of its request surface; each route in that graph is then audited on its own by \emph{invariant falsification}, where an agent enumerates the properties the route must guarantee and \textit{constructs the request that would violate each}; a final stage links findings and capabilities across routes.}
    \label{fig:pipeline}
\end{figure}

Three requirements follow. \textbf{R1:} The analysis must recover the application's request surface before it judges any of it, since whatever enforces a route need not sit near it, and applications differ in way they declare routes and attach guards. \textbf{R2:} It must then proceed over that surface route by route, so that what is read is settled in advance rather than by what draws an agent's attention. \textbf{R3:} It must establish the property a route is meant to guarantee before checking the code against it, and report a finding as the request that would violate that property rather than as a judgment, since no rule supplies it in advance and a model asked only whether code looks wrong has no standard to answer against. \S\ref{sec:design} presents the scaffolding that meets them (Figure~\ref{fig:pipeline}), together with a stage that composes findings across routes and one that reuses verdicts a later commit cannot have changed.

\section{Design}
\label{sec:design}

\method takes the source of one application and returns findings, each naming a property that one route is meant to guarantee together with the request that would violate it. This section presents the scaffolding that produces them. \S\ref{sec:design-shared-fact-substrate} builds a graph of the application's request surface, so that what will be analyzed is settled before any of its in-depth code reading. \S\ref{sec:design-invariant-falsification} then analyzes each route in that graph on its own, and fixes what a finding has to contain. \S\ref{sec:design-chaining} links findings across routes, for violations that no single route reveals. \S\ref{sec:design-incremental} reuses the verdicts that a later commit cannot affect, since an application under development is analyzed more than once.

\subsection{Mapping the Application}
\label{sec:design-shared-fact-substrate}

Given only an application's source code, \method must find the access-control failures of \S\ref{sec:scope}. Real applications make that a search through a great deal of code: the backends in our benchmark hold a median of 151 thousand lines and the largest more than a million (\S\ref{sec:benchmark}), of which the code that serves requests is only a small part. \method has to find that part before it can judge any of it.
\par
A security engineer auditing the same application does not read files at random \cite{owasp-wstg-42}. They first orient themselves, working out what the application is built on, because that is what tells them where the code worth reading lives. Only then do they work through the application's surface, route by route, asking of each what it exposes and what stands in front of it.
\par
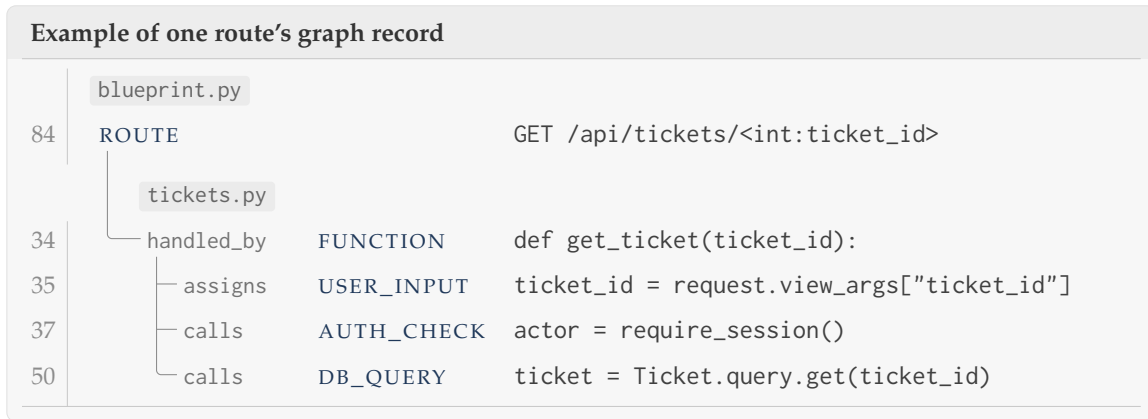
\begin{figure}[t]
  \centering
\begingroup
\begin{tikzpicture}[
    x=1mm,y=-1mm,
    gut/.style={anchor=base east,inner sep=0pt,
      font=\fontsize{9}{11}\selectfont\rmfamily,text=black!45},
    badge/.style={anchor=base west,inner sep=0pt,
      font=\fontsize{9.4}{11}\selectfont\rmfamily\scshape,text=nodeink},
    elab/.style={anchor=base west,inner sep=0pt,
      font=\fontsize{9}{11}\selectfont\ttfamily\frenchspacing,text=black!55},
    src/.style={anchor=base west,inner sep=0pt,
      font=\fontsize{10}{12}\selectfont\ttfamily\frenchspacing,text=black!75},
    fhd/.style={anchor=base west,
      font=\fontsize{9}{11}\selectfont\ttfamily\frenchspacing,text=black!60,
      fill=black!8,rounded corners=.55mm,inner xsep=1.1mm,inner ysep=.65mm},
    hdr/.style={anchor=base west,inner sep=0pt,
      font=\fontsize{9.5}{11}\selectfont\rmfamily\bfseries,text=black!80},
    rule/.style={draw=black!22,line width=.3pt},
    tree/.style={draw=black!45,line width=.4pt,rounded corners=.6mm}
  ]
  \def\xgut{6.5}
  \def\xbar{8}
  \def\xbadge{41}
  \def\xsource{67}
  \def\xend{152}
  \newcommand{\recordrow}[6]{%
    \node[gut] at (\xgut,#2) {#1};
    \node[elab] at (#3,#2) {#4};
    \node[badge] at (\xbadge,#2) {#5};
    \node[src] at (\xsource,#2) {#6};
  }

  \node[hdr] at (3,4.7) {Example of one route's graph record};
  \draw[rule] (2,7) -- (\xend-2,7);

  \node[fhd] at (10.9,12) {blueprint.py};
  \node[gut] at (\xgut,18) {84};
  \node[badge] at (12,18) {route};
  \node[src] at (\xsource,18) {GET /api/tickets/<int:ticket\_id>};

  \node[fhd] at (17.4,26) {tickets.py};
  \recordrow{34}{32}{18.5}{handled\_by}{function}{def get\_ticket(ticket\_id):}
  \recordrow{35}{38}{23.3}{assigns}{user\_input}{ticket\_id = request.view\_args["ticket\_id"]}
  \recordrow{37}{44}{23.3}{calls}{auth\_check}{actor = require\_session()}
  \recordrow{50}{50}{23.3}{calls}{db\_query}{ticket = Ticket.query.get(ticket\_id)}

  \draw[tree] (13.3,19.2) -- (13.3,30.8) -- (17.7,30.8);
  \draw[tree] (19.8,33.2) -- (19.8,48.8) -- (22.5,48.8);
  \foreach \row in {38,44}
    \draw[tree] (19.8,\row-1.2) -- (22.5,\row-1.2);

  \draw[rule] (\xbar,8.3) -- (\xbar,21);
  \draw[rule] (\xbar,28.5) -- (\xbar,53);
  \draw[rule] (2,53) -- (\xend-2,53);

  \begin{pgfonlayer}{background}
    \fill[black!3,rounded corners=1.2mm] (0,0) rectangle (\xend,55);
    \fill[black!7,rounded corners=1.2mm] (0,0) rectangle (\xend,7);
    \fill[black!7] (0,5) rectangle (\xend,7);
    \draw[black!12,line width=.3pt,rounded corners=1.2mm]
      (0,0) rectangle (\xend,55);
  \end{pgfonlayer}
\end{tikzpicture}
\endgroup
  \captionsetup{width=.93\linewidth}
  \caption{Illustrative record of one graph route: each row a node with its type and line, each connector a typed edge.}
  \label{fig:route-record}
\end{figure}

\method follows the same order. The surface has to be found before it can be worked through, so \method first maps it into a graph~\cite{locagent}. That graph records where each route is declared, the code that serves it, the authorization checks the application declares, the request values that code reads, and the sensitive operations behind them, such as the database queries and the writes into the response. Each of these is recorded as a file and a line, and the graph records which of them belong to the same route. Figure~\ref{fig:route-record} shows an example of what the graph could hold for one route, whose elements were all declared in the same file. The graph settles what will be analyzed and where to look, and judges none of it; judgment happens one route at a time (\S\ref{sec:design-invariant-falsification}).

\par

However, two applications rarely declare their routes in the same way. A framework sets the broad conventions, the version narrows them, and the application then wraps them in helpers and base classes of its own, so the expressions precise enough to locate a route, a request value, a query or a guard are usually particular to one application. \method therefore has an LLM write them, since a model can recognize the conventions it has seen during training and can also work out unfamiliar ones from the code. Two passes follow: profiling establishes what the application is built on, and extraction produces the expressions that locate each kind of element.

\myparagraph{Profiling.} An agent lists the directory layout, opens the dependency manifests, and searches for the imports that show which of the declared dependencies the code actually uses. It reports a short list of labels: the framework and language, the data stores and the libraries the application queries them through, and the libraries by which it authenticates its callers. For Indico, a conference-management application among the 25 of \S\ref{sec:benchmark}, that list reads Flask over Python, PostgreSQL and SQLite through SQLAlchemy, and sessions and OAuth through Flask-Multipass.\footnote{\url{https://github.com/indico/indico/}} The same report answers one further question: how this application decides who may see which records. Some applications separate callers by ownership of a record, others by ranked roles, and others by both. That choice is a property of the application rather than of its framework. The agent has to work it out from the code. Without that answer, a handler that returns the record named in a request cannot be judged correct or incorrect.

\myparagraph{Extraction.} Obtaining these profiling labels is worth a pass of its own because they narrow what the extraction agents have to look for. Knowing that the application is Flask does not say which expressions declare its routes, but it does say which family of expressions to expect, and the same holds for a query issued through SQLAlchemy or a caller identified through Flask-Multipass. An extraction agent therefore starts from a known framework, data layer and authentication library, and has only to recover the expressions this application reaches each one through.
\par
With that report, the application is read again, this time in depth. An agent opens its routing and handler files and reports, for each kind of element, which expression marks it and which files to search for that expression. For Indico the answers come back as the expressions particular to it: \texttt{\_bp.add\_url\_rule} for a route declaration, \texttt{request.view\_args} for a request value, \texttt{db.session} for a query, \texttt{session.user} for the caller. Two agents do this independently and their reports are unioned.
\par
\method applies the crafted patterns deterministically, and every match becomes an element of the graph once duplicates and false matches are dropped. A route no pattern matches is absent from the graph and from every stage after it, which is why the two reports are unioned rather than either taken alone. Because one declaration can serve several HTTP verbs, a further pass reads the source around each route to settle its method and path, and expands it into one route per method, so that the surface is a set of routes rather than of registration sites.

\subsection{Per-Route Invariant Falsification}
\label{sec:design-invariant-falsification}

\myparagraph{Per-route decomposition.} By default, \method analyzes every route in the graph. A handler that looks harmless can still be the one missing a check, so it is safer to read every route once than to leave one out. Reading every route is also where most of the cost of running \method falls, and \S\ref{sec:eval:rq4} reports that cost together with the share of route verdicts that carry over when a repository is re-analyzed at a later commit.
\par
An agent receives the graph's record of its route (Figure~\ref{fig:route-record}), which names the registration site and the code that the graph linked to it, and it opens that code itself. It reads the handler first and then follows the calls out of it, into the services and data models behind it, as far as it takes to see where each input is consumed and where each response field is produced. The rest of the application stays outside its context unless it goes looking, so how much code one agent holds at once is set by the route it was given rather than by the size of the application around it.

\myparagraph{Invariant falsification.} Everything so far settles what to read; this step decides whether a defect is found. The agent takes one route at a time and works through it in a fixed order. It first enumerates the properties the route is meant to guarantee. For each, it then locates the code that is supposed to enforce it. Finally, it tries to falsify each by describing a request that the endpoint would accept and that the property forbids.
\par
The properties are the agent's own. It states each in plain language, after reading the handler and the data model behind it, as a claim about this route in this application. Figure~\ref{fig:pipeline} shows one: an invoice is returned only to the account that owns it, which holds in the application of \S\ref{sec:overview} and would not hold in one where invoices are shared. Recovering the properties a program is expected to satisfy is the aim of Daikon~\cite{daikon} and of work applying it to logic flaws in web applications~\cite{logic-vulns-usenix-2010}. Both recover them from executions of the program, but \method infers them from its source.
\par
An enforced property produces no report. A missing or insufficient enforcement produces a finding with a fixed schema: the candidate invariant, the relevant code locations, the request shape, the preconditions under which it applies, and the effect that would demonstrate the violation (Figure~\ref{fig:pipeline}). The agent can write those only from code it has read, so the form gives it nowhere to put a suspicion it cannot ground in source, and the precision this buys is what \S\ref{sec:eval:rq2} measures.

\subsection{Composition Across Routes}
\label{sec:design-chaining}

Findings are not always bound to a single route: what is minor on one route can be serious once a second route is in play. \method therefore has each agent record, besides its findings, the behaviors of its route that break no invariant but would matter if a second route did something in particular: a capability, written down as the behavior, what the caller controls in it, and what would have to hold elsewhere. Figure~\ref{fig:pipeline} shows the shape of it. An upload that stores a file under a name the caller chooses is a capability, harmless while nothing serves that store back to callers. A second route that does serve files out of it, without checking what they are, is a small matter on its own, since that route can assume the store holds what the application itself put there. Together they let a caller control what the application serves to anyone who asks, and an agent reading either route alone has no cause to report it.
\par
Pairs like that are what \method looks for once every route has been analyzed. Two routes declared in different places still meet in the code they act on, so \method matches the findings and capabilities located in the same file, and gives each pair to an agent that reads both routes and that shared code before it reports the composition. \S\ref{sec:discussion-limitations} reports how far this stage was exercised on \bench.

\subsection{Re-Analyzing After a Commit}
\label{sec:design-incremental}

An audit is not a single event. An application under development has to be checked repeatedly, so that a defect is found before it ships rather than after, which puts \method on a sequence of commits rather than on one snapshot. Analyzing every route in the graph is what makes its coverage systematic (\S\ref{sec:design-invariant-falsification}), but also what a run spends its budget on. \method therefore decides, between two commits, which of the previous run's verdicts still stand and can be reused.
\par
That decision reads each commit's graph together with the files the diff between them changed. Since a graph names the files every route is built from, a verdict from the earlier commit comes with the set of files it was read from, widened along the imports those files declare so that code the route reaches through another file counts as its own. \method keeps that verdict when no file in that set appears in the diff and all of them are byte-identical across the two commits, and analyzes the route again otherwise.

\section{\bench}
\label{sec:benchmark}

\bench is a benchmark for broken access control detection from source code. It holds 30 disclosed authorization advisories, each validated by a human auditor, drawn from 25 open-source projects. Each instance is the code of one project at the commit where the disclosed defect was live, paired with the commit that fixed it. 

\begin{table}[htbp]
  \centering
  \caption{Composition of \bench. Each instance is one disclosed advisory, counted under the first weakness (CWE) class its advisory assigns.}
  \label{tab:dataset}
  \small

  \setlength{\tabcolsep}{3.5pt}
  \begin{minipage}[t]{.49\linewidth}
  \vspace{0pt}
  \begin{tabularx}{\linewidth}{@{}>{\raggedright\arraybackslash}X r@{}}
    \toprule
    \textbf{Dimension} & \shortstack[r]{\textbf{Count /}\\\textbf{statistic}} \\
    \midrule
\multicolumn{2}{@{}l}{\emph{Scale}} \\
    Disclosed advisories (instances)  & 30 \\
    Repositories                      & 25 \\
    Web frameworks                    &  9 \\
    \addlinespace
\multicolumn{2}{@{}l}{%
      \emph{Repository characteristics (median [min--max])}} \\
    Backend size (LoC)                & 151k [12k--1.2M] \\
    Routes                            & 220 [17--999] \\
    GitHub stars                      & 8.5k [91--199k] \\
    \bottomrule
  \end{tabularx}
  \end{minipage}\hfill
  \begin{minipage}[t]{.47\linewidth}
  \vspace{0pt}
  \begin{tabularx}{\linewidth}{@{}>{\raggedright\arraybackslash}X r@{}}
    \toprule
    \textbf{Dimension} & \shortstack[r]{\textbf{Count /}\\\textbf{statistic}} \\
    \midrule
\multicolumn{2}{@{}l}{\emph{Primary CWE (instances)}} \\
    CWE-639 (IDOR/BOLA)               & 15 \\
    CWE-862 (missing authorization)   & 10 \\
    Other (CWE-284, CWE-285, CWE-863) &  5 \\
    \addlinespace
\multicolumn{2}{@{}l}{\emph{Implementation language (instances)}} \\
    Python                            & 18 \\
    TypeScript                        &  9 \\
    JavaScript                        &  3 \\
    \bottomrule
  \end{tabularx}
  \end{minipage}
\end{table}

\subsection{Construction}
\label{sec:benchmark-construction}
\myparagraph{Selection.}
An advisory becomes an instance when it concerns a web application written in Python, JavaScript or TypeScript, and five conditions are guaranteed: (i) the advisory describes a genuine authorization defect, such as CWE-639 or CWE-862; (ii) a vulnerable commit exists at which the defect is live; (iii) a commit exists whose diff is the upstream fix for that defect; (iv) the advisory was published in 2025 or later; and (v) a security researcher has validated the instance against the advisory and the fix diff, correcting every field where the public record and the code disagreed.

\myparagraph{The oracle.}
The oracle for an instance is the advisory together with the vulnerable code and the fix diff. It records the invariant the route was meant to guarantee, the way the code breaks it, the routes affected, and two sets of locations. The first is the files named in the advisory and in the pull request that fixed it. The second is the lines the fix changed, mapped back to the range they occupied in the vulnerable code. \S\ref{sec:matching} uses the second to separate a finding that points at the defect from one that points only at its neighborhood.

\myparagraph{Composition.}
Table~\ref{tab:dataset} reports how instances are distributed over weakness class, language and scale. The 9 frameworks span four Python stacks and five in the JavaScript family. Both shapes of failure described in \S\ref{sec:two-families} occur: in some instances the fix adds a check where none stood, and in others a check was present but bound the wrong principal, object or scope.
\par
Each instance names the one bug its advisory identifies, and we ask only whether a tool reported that bug. At the commit that fixed it, we expect no tool to report that bug. Both questions are narrow by design. A large application may contain several authorization bugs, a single commit may contain more than one, and some are never disclosed, so we do not know every defect a repository contains and do not ask a tool to enumerate them. The negative is therefore specific to the bug under test, since the fixed version still holds whatever nobody has disclosed. Findings that name other weaknesses in the same repository are neither credited nor penalized, and precision is measured separately in \S\ref{sec:evaluation}.

\subsection{Adjudication}
\label{sec:matching}

Deciding whether an instance was detected must not bias the evidence format of any one tool. A static analyzer reports a file, a line, a rule identifier and a short message~\cite{semgrep,codeql}, where an agent may additionally report more detailed outputs such as a rationale in prose. Findings also arrive in volume, so the procedure has to settle many of them with as little human effort as possible. We therefore adjudicate every tool through the same two stages.

\myparagraph{The location gate.}
The first stage is deterministic, and its goal is to filter out the findings that do not concern the bug under focus. A finding survives when its location overlaps the instance's implicated files or when its route matches an affected route. For example, across the vulnerable commits of all 30 instances Semgrep reports 6,398 findings, of which only 66 move forward.

\myparagraph{The judge.} The second stage is where a detection is decided. The gate leaves few candidates per instance, so we can afford a language model as judge (Claude Sonnet 4.6 \cite{claude-4-6-Sonnet}) that reads each one against the invariant and the way the code breaks it. It is not told which tool produced the finding, and it runs at temperature zero. It also reads no repository source, so it adjudicates the claim a tool made rather than re-analyzing the code on that tool's behalf.

\myparagraph{Validating the judge.}\tinyquad Every recall number in this paper rests on a model's judgment of what counts as a match, so we measured that judgment against a person's. A graduate researcher in security replayed the judge's task on every match it credited and a sample of its rejections, given the same oracle and the same finding text and shown neither the judge's decision nor which tool produced the finding: the two agree on \result{81} of \result{85} adjudications, Cohen's $\kappa$ of \result{0.91}.

\section{Implementation}
\label{sec:implementation}

\method runs each stage of Figure~\ref{fig:pipeline} as a Strands agent~\cite{strands} over MiniMax-2.5~\cite{minimax}. We choose this pairing for the setting of \S\ref{sec:scope}, where the code under audit is often private and an organization may be unwilling to send it to a third-party service. Strands is open source and drives any OpenAI-compatible endpoint, and MiniMax-2.5 is open-weight, so the whole pipeline can run inside that organization, on its own machines, under a local server such as vLLM \cite{vllm}. We use AWS Bedrock\footnote{\url{https://aws.amazon.com/bedrock/}} in our experiments, but the method does not depend on it. The costs \S\ref{sec:eval:rq4} reports are the tokens runs consumed at the rates that service charged for this model, \$0.30 per million input tokens and \$1.20 per million output. Three further models appear in the evaluation, none of them inside \method: Claude Sonnet 4.6 \cite{claude-4-6-Sonnet} adjudicates detection (\S\ref{sec:matching}), Claude Opus 4.8 \cite{claude-4-8-Opus} verifies precision and Claude Sonnet-5 \cite{claude-5-Sonnet} stands in as the stronger reader (\S\ref{sec:eval:rq2}).
\par
Temperature is zero wherever a stage reasons about an individual route, though not in the stages that build the graph (\S\ref{sec:design-shared-fact-substrate}): extraction deliberately runs two agents at 0.1 and 0.7 and unions their reports, so that a route one pass misses can still be recovered by the other.
\par
All access to the application is read-only and confined to the repository directory: reading a slice of a file, listing the source files under a directory, searching the sources for a regular expression, and resolving the imports a file declares. \method runs no part of the application which is what \S\ref{sec:scope} means by an audit of source code.
\section{Evaluation}
\label{sec:evaluation}

We first state the datasets, the baselines and the metrics (\S\ref{sec:setup}), then report \method's detections on the benchmark of \S\ref{sec:benchmark}, the contribution of each part of the scaffolding to that result, whether the same configuration reaches beyond authorization, and the cost of a run together with its amortization across commits (\S\ref{sec:evaluation-rqs}).

\subsection{Experimental Setup}
\label{sec:setup}

\myparagraph{Datasets.}
The evaluation spans two families of datasets. The first carries the main result: \bench, 30 disclosed broken-access-control advisories across 25 real Python, TypeScript and JavaScript web applications, with the patched-commit differential described in \S\ref{sec:benchmark}.

\par

The second family tests whether \method reaches beyond authorization: the OWASP Benchmarks in Python and in Java~\cite{owasp-benchmark-python,owasp-benchmark-java}. Each is a synthetic application labeled at construction, safe cases included, which lets us report a false-positive rate here and not on the authorization data. Their categories are broader than the vulnerabilities a source-to-sink analysis is meant to address, since they also cover classes such as weak randomness, insecure cookie flags and broken hashing. We therefore evaluate both \method and the baselines on the injection subset of each benchmark, which leaves 10 categories in Python and 6 in Java, among them SQL and command injection, cross-site scripting and path traversal. In total, there are 677 cases in Python, of which 240 are vulnerable, and 1{,}572 in Java, of which 819 are vulnerable.

\myparagraph{Baselines.}

We compare \method against five baselines, covering the three ways static analysis can reach these defects today: rules written in advance, a language model on its own, and the two combined. Two are standard static analyzers: CodeQL with its security-extended query suites, and Semgrep with the security, OWASP and language packs. Both are deterministic and have no language model. The third is IRIS~\cite{iris}, in which a language model infers a repository's taint source and sink specifications, those compile into CodeQL taint queries, and a second model pass filters the results. IRIS was released for Java only, so we keep the published implementation and re-implement the same method faithfully for Python. The fourth is RepoAudit \cite{repoaudit}, an agentic data-flow analyzer that explores a repository and validates candidate paths with a model rather than compiling queries. The fifth, \vanilla, is the unstructured control, asked to audit the repository end to end for 200 turns but without a route graph or per-route invariant falsification. To isolate the analysis from the model, all three model-driven baselines run on the model \method uses.
\par
Baseline coverage differs by problem type. CodeQL, Semgrep and \vanilla run on all 30 authorization instances. \vanilla is the unstructured control and has to match \method's coverage. CodeQL and Semgrep are deterministic and free to run, so we sweep them in full even though we expect no detections. IRIS and RepoAudit run on injection only. Both are model-driven and priced per repository, and a sweep of either would return zeros set by what the tool can state, not by what the benchmark holds. RepoAudit ships no JavaScript or TypeScript analyzer, which puts 12 of the 30 instances out of reach, and its engine asks only whether a source reaches a sink. IRIS instantiates its method once per taint class and ships no authorization query; adding one would make the baseline our construction rather than the published method. The model budget goes to injection instead, where a flow is the right abstraction.

\myparagraph{Metrics.}

Recall is the primary metric on \bench because its targets are real applications, where the vulnerabilities nobody has disclosed are unknown and the safe code therefore cannot be enumerated. On top of recall on the vulnerable commit we add two things: (i) a controlled negative, where each tool runs on the fixed commit of every instance as well, so that a finding persisting where the disclosed defect no longer exists is adjudicated as a false positive; and (ii) precision by verification, the share of a tool's emitted findings that an independent verifier confirms against the code, which \S\ref{sec:eval:rq2} reports together with a blinded human check of that verifier.
\par
Because \method is nondeterministic (\S\ref{sec:implementation}), a single run does not settle its recall, so we repeat it and report three quantities, each averaged over instances: the recall of one run (pass@1), the recall counting an instance detected if any of the $k$ runs detects it (pass@$k$), and the recall counting it detected only if all $k$ do (pass$^k$ \cite{tao-bench}). We run this experiment on the eight instances below 100 routes with $k{=}5$.
\par
The injection suites have safe cases by construction, so we can report a false-positive rate on them. We adopt the OWASP Benchmark's own scoring plugin and the metrics it recommends: the true-positive rate (TPR), the false-positive rate (FPR), and their difference, Youden's~J. We macro-average all three over the injection categories of each suite, so that a category with many cases does not outweigh one with few.

\subsection{Research Questions}
\label{sec:evaluation-rqs}

\noindent\method's evaluation answers four main questions:
\begin{description}
  \item \textbf{RQ1 (Broken access control detection).} How many disclosed BAC vulnerabilities does \method detect, does it stay silent on the commits that fixed them, and where does the recall it does not reach go?
  \item \textbf{RQ2 (Component Contributions).} What does each part of the method contribute: the per-route decomposition, the invariant falsification, and the model underneath them? And is the precision we report an artifact of adjudicating with a model?
  \item \textbf{RQ3 (Generalization Beyond Authorization).} Does the same configuration hold up against dedicated static analyzers and model-driven taint tools on injection, in Python and in Java, or is the authorization result bought by specializing to that one class?
  \item \textbf{RQ4 (Cost and Continuous Analysis).} What does a run cost, what governs that cost, and how much of it is avoided when a repository is re-analyzed after a commit?
\end{description}

\FloatBarrier
\subsection{Broken Access Control Detection (RQ1)}
\label{sec:eval:rq1}

\begin{table}[htbp]
  \centering
  \captionsetup{width=.85\linewidth}
  \caption{Recall on \bench (\result{30} disclosed advisories). We report recall on the vulnerable commit and the paired recall that credits an instance only when it is detected on the vulnerable commit and cleared on the patched commit.}
  \label{tab:authz-recall}

\begin{tabularx}{.85\linewidth}{@{}X c c@{}}
  \toprule
  \textbf{Tool}
    & Recall$_{\text{vuln}}\!\uparrow$
    & Paired$\ \!\uparrow$ \\
  \midrule
  CodeQL   & 0/30  & 0/30 \\
  Semgrep  & 0/30  & 0/30 \\
  \vanilla & 3/30  & 3/30 \\
  \method  & \textbf{19/30} & \textbf{17/30} \\
  \bottomrule
\end{tabularx}
\end{table}

\myparagraph{Detection recall.} \method detects 19 of the 30 disclosed vulnerabilities, the two static analyzers detect none of them, and \vanilla detects 3 (Table~\ref{tab:authz-recall}). The analyzers are not short of reach: they scan the whole repository, so the code holding each defect is read on every instance, and across the vulnerable commits Semgrep alone emits 6{,}398 findings. They just end up being unhelpful because their query set, while vast, is generic, and the properties of authorization are specific to each of the applications.
\par
\vanilla, on the other hand, is affected by another issue, ineffective exploration. It was given 200 turns to analyze the repository, but on average it used 59, ending on 24 of the 30 runs by declaring itself finished, and on a full context window on the other 6. Its 3 detections therefore do not reflect a budget we imposed. The same model detects 19 of the 30 when the route graph directs it to analyze each route separately.
\par
However, detecting the defect is only half of what the benchmark asks. A report that persists after the fix tells a reviewer nothing, so the paired column requires both: present at the vulnerable commit, absent at the one that fixed it. \method holds on 17 of its 19, and on the remaining 2 it reports the defect at the commit that fixed it, which are cases where it reported the bug incorrectly.

\begin{figure}[htbp]
    \centering
    \includegraphics[width=0.76\linewidth]{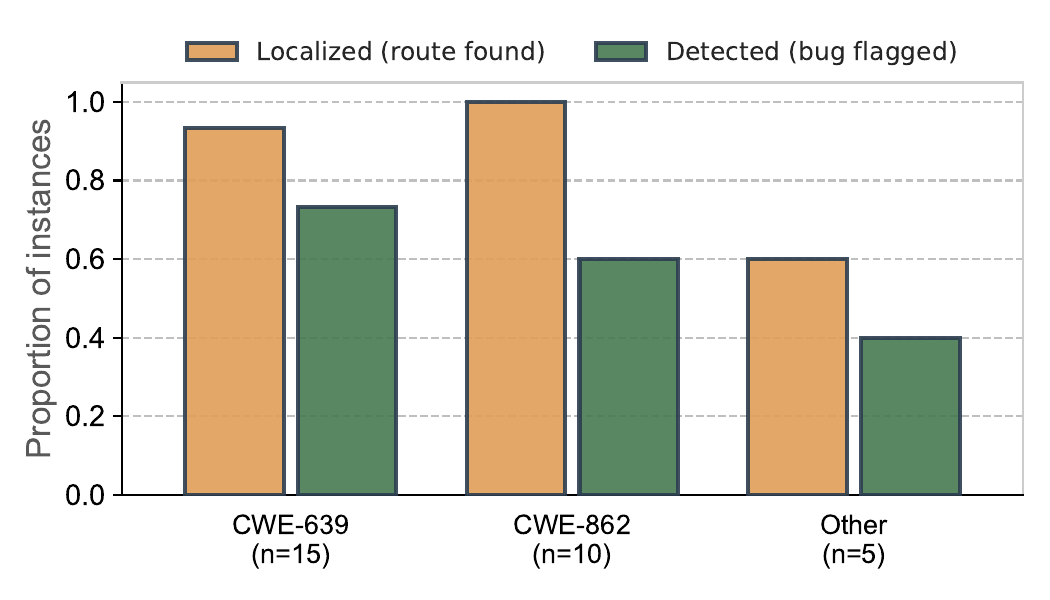}
    \captionsetup{width=.82\linewidth}
    \caption{Where \method's recall goes, by weakness class. Nearly every vulnerable route reaches the reasoning stage, so most of what \method misses is lost after the route is found rather than in finding it.}
    \label{fig:funnel}
\end{figure}

\myparagraph{Localization and reasoning.} \method reaches an instance through two gates in sequence, and Figure~\ref{fig:funnel} separates them. The vulnerable route has to be in the graph for any agent to be given it, and the agent then has to falsify the invariant the route breaks. Of the 30 instances, 27 clear the first gate and 19 clear the second, so extraction accounts for 3 of the 11 misses and reasoning for the remaining 8. The two failures are not equivalent. A route absent from the graph is never read by anything, and the defect cannot be reported at any level of reasoning skill; a route that is present and read wrongly is a judgment we can improve on.
\par
Read by class, no group carries the result. In both large groups the drop at the reasoning gate is larger than the drop at extraction, so the limit is the reasoning stage rather than the weakness class. The 3 routes never recovered come from 3 different frameworks with no cause common to them, and we therefore treat them as individual gaps rather than as a surface the graph systematically fails on.

\myparagraph{Run-to-run stability.} Extraction runs two agents at different temperatures (\S\ref{sec:implementation}), so two runs over the same commit will not necessarily report the same findings, and one run does not settle whether \method detects an instance. Table~\ref{tab:reliability} repeats the eight instances below 100 routes five times each, 40 full runs in all.

Six of the eight return the same verdict every time, four detected and two missed, so an instance is mostly detected or missed as a property of itself. What varies is concentrated in two instances, aegra at 4 of 5 trials and rallly at 3 of 5, so the effect on the aggregate is bounded: recall is 0.68 for a single run, 0.75 if one of the five suffices, and 0.50 if all five must agree.
\par

\begin{table}[htbp]
  \centering
  \captionsetup{width=.82\linewidth}
  \caption{Reliability of \method\ under nondeterminism on the instances below \result{100} routes, each run $k{=}5$ times, as pass@1 (mean detection over trials), pass@$k$ (detected in at least one) and pass$^k$ (detected in every one). Six of eight yield identical verdicts across runs, suggesting detection is largely a property of the instance rather than the run.}
  \label{tab:reliability}
  \small
    \begin{tabular*}{.82\linewidth}{@{\extracolsep{\fill}}l r c c c@{}}
      \toprule
      \textbf{Instance} & Routes & pass@1 & pass@$k$ & pass$^k$ \\
      \midrule
      payload-puck  &  17 & 1.00 & 1.00 & 1.00 \\
      git-proxy     &  30 & 0.00 & 0.00 & 0.00 \\
      aegra         &  46 & 0.80 & 1.00 & 0.00 \\
      actualbudget  &  51 & 1.00 & 1.00 & 1.00 \\
      FUXA          &  59 & 1.00 & 1.00 & 1.00 \\
      parse-server  &  89 & 0.00 & 0.00 & 0.00 \\
      rallly        &  95 & 0.60 & 1.00 & 0.00 \\
      khoj          &  98 & 1.00 & 1.00 & 1.00 \\
      \midrule
      \textbf{Average} &  & 0.68 & 0.75 & 0.50 \\
      \bottomrule
    \end{tabular*}
\end{table}
Two limits on this reading are worth stating. The subset is the small end of the benchmark, chosen because five trials on the 22 larger instances would cost more than the rest of the evaluation together, and those instances are run once; nothing here establishes that a 600-route repository is as stable as a 100-route one. And at $k{=}5$ a per-instance rate moves in steps of $0.2$, so the table characterizes run-to-run stability rather than estimating a per-instance detection probability. Raising $k$ can only expose more variation, which is what happened here: at $k{=}3$ aegra was unanimous, and the two additional trials moved pass$^k$ from 0.62 down to 0.50.

\FloatBarrier
\subsection{Component Contributions (RQ2)}
\label{sec:eval:rq2}

Three things could be doing the work in \method: the route graph that decides what an agent is given to read, the falsification step that decides what becomes a finding, and the model that reads the code. We separate them in that order and then ask whether the precision they are measured by holds up when a person checks the verifier against the same code.

\begin{table}[htbp]
  \centering
  \caption{Capability spectrum from free exploration to the full method on vulnerable commits, with each row adding one capability. Per-route decomposition drives the recall gain, from 3/30 to 18/30; invariant falsification then cuts reports by a third and raises verified precision from 40.5\% to 50.8\%, at a one-instance difference in recall that is inside run-to-run variation (Table~\ref{tab:reliability}).}

  \label{tab:ablation}
  \small
  \begin{tabular*}{\linewidth}{@{\extracolsep{\fill}}l c c c@{}}
    \toprule
    \textbf{Configuration} & Recall & Avg.\ \#reports & Precision \\
    \midrule
    \vanilla         & 3/30 & 6 & 73.6\% \\
    $+$ per-route decomposition             & 18/30 & 119 & 40.5\% \\
    $+$ invariant falsification    & 19/30 & 77 & 50.8\% \\
    \bottomrule
  \end{tabular*}
\end{table}

\myparagraph{The two components.}  Table~\ref{tab:ablation} takes the method apart in the order the design introduces it. The middle row keeps the route graph and the per-route decomposition but asks each agent only for a weakness class and a rationale, without the invariant, the enforcing code or the falsifying request. The bottom row is \method as \S\ref{sec:design} describes it.
\par
Decomposition is what makes the vulnerabilities reachable, from 3 instances to 18, and adding falsification on top of it moves recall by one. What falsification does change is the volume: the average instance yields 77 reports rather than 119, a third fewer, with the disclosed defects still found. That much needs no judgment of report quality at all.
\par
\myparagraph{Verified precision.}  Falsification suppresses 42 reports per instance on average, and whether those were the right ones to suppress is a question of precision. Precision has no oracle here: these are production applications, and nothing certifies the code a report points at as safe unless that report is the disclosed defect. So we verify each report against the code instead. An independent agent using Claude Opus 4.8 \cite{claude-4-8-Opus} reads the repository and decides whether the code substantiates each report, over every instance, capped at 20 reports drawn at random from each configuration and never told which one wrote them. In its reading, the confirmed share rises from 40.5\% to 50.8\%, a paired gain of 10.3 points.
\par
Since that verdict comes from a model, a security reviewer audited 100 of its decisions blind, reading the same code without seeing the verifier's answer. On whether a finding is substantiated at all, the two broadly agree. They return the same verdict on 85\% of the 73 findings both read as authorization claims, with Cohen's $\kappa$ of 0.70, and per configuration that agreement is 82\% on \method and 89\% on the other (Table~\ref{tab:verifier-human}). Where they part is the gap itself: the reviewer confirms \method at 42\% and the other configuration at 43\%, while the verifier separates them at 50\% and 43\%.
\par

\begin{table}[htbp]
  \centering
  \captionsetup{width=.82\linewidth}
  \caption{The precision verifier against a blinded human reviewer, over the 73
  findings both read as authorization claims. The reviewer confirms the two
  configurations at the same rate where the verifier separates them, so this
  sample does not reproduce the gap of Table~\ref{tab:ablation}.}
  \label{tab:verifier-human}
  \small
  \begin{tabular*}{.82\linewidth}{@{\extracolsep{\fill}}l c c@{}}
    \toprule
    \textbf{Measure} & \method & no falsification \\
    \midrule
    Agreement with verifier    & 82\% & 89\% \\
    Confirmed by human expert      & 42\% & 43\% \\
    Confirmed by verifier      & 50\% & 43\% \\
    \bottomrule
  \end{tabular*}
\end{table}

As a consequence, we cannot claim that the verifier's gap is adjudicated by the expert. Those two rates are a point apart where the verifier's are seven. We therefore report the precision gain as the verifier measures it, but not as a quantity the expert could confirm at this sample size. What does not depend on it is the volume: falsification reports a third fewer findings than the configuration without it and loses no recall doing so. As such, what it discards is not where the disclosed defects were.

\myparagraph{The model underneath.} Of the 11 vulnerabilities \method does not report, Figure~\ref{fig:funnel} placed 8 after the vulnerable route was localized, where an agent read the route and did not report the defect, and 3 before it, where the route never entered the graph. Neither number says how much of either loss belongs to the model doing the reading. We therefore repeat both stages with Claude Sonnet-5 \cite{claude-5-Sonnet}, a more capable model, in place of MiniMax-2.5 (Table~\ref{tab:frontier}).

\begin{table}[htbp]
  \centering
  \captionsetup{width=.82\linewidth}
  \caption{Effect of a stronger model on each stage, scored over the 27 instances whose vulnerable route was localized and the 3 where it was not. Even handed the route, a stronger model recovers only a few more instances and none of the routes extraction never produced, so the routes the graph recovers are what a stronger reader can act on.}
  \label{tab:frontier}
  \small
  \begin{tabular*}{.82\linewidth}{@{\extracolsep{\fill}}l c c@{}}
    \toprule
    \textbf{Stage} & MiniMax-2.5 & Sonnet-5 \\
    \midrule
    Route with bug extracted & 27/30 & 27/30 \\
    Bug reported & 19/27 & 22/27 \\
    \bottomrule
  \end{tabular*}
\end{table}

On extracting the route the swap changes nothing: re-run on the 3 instances whose vulnerable route never entered the graph, Sonnet-5 recovers none of them. On reporting the bug it does move. Here Sonnet-5 reads MiniMax-2.5's own extracted routes so that per-route reasoning is the only thing that varies: it reports 22 of those 27 instances where MiniMax-2.5 reports 19. That gain is real but not clean, because the two sets cross rather than nest, with Sonnet-5 losing an instance MiniMax-2.5 reports. A movement of this size is also within what repeated runs of one configuration produce on their own (Table~\ref{tab:reliability}).

\FloatBarrier
\subsection{Generalization Beyond Access Control (RQ3)}
\label{sec:eval:rq3}

\begin{figure}[tbp]
    \centering

    \begin{subfigure}[t]{0.49\linewidth}
        \centering
        \includegraphics[width=\linewidth]
            {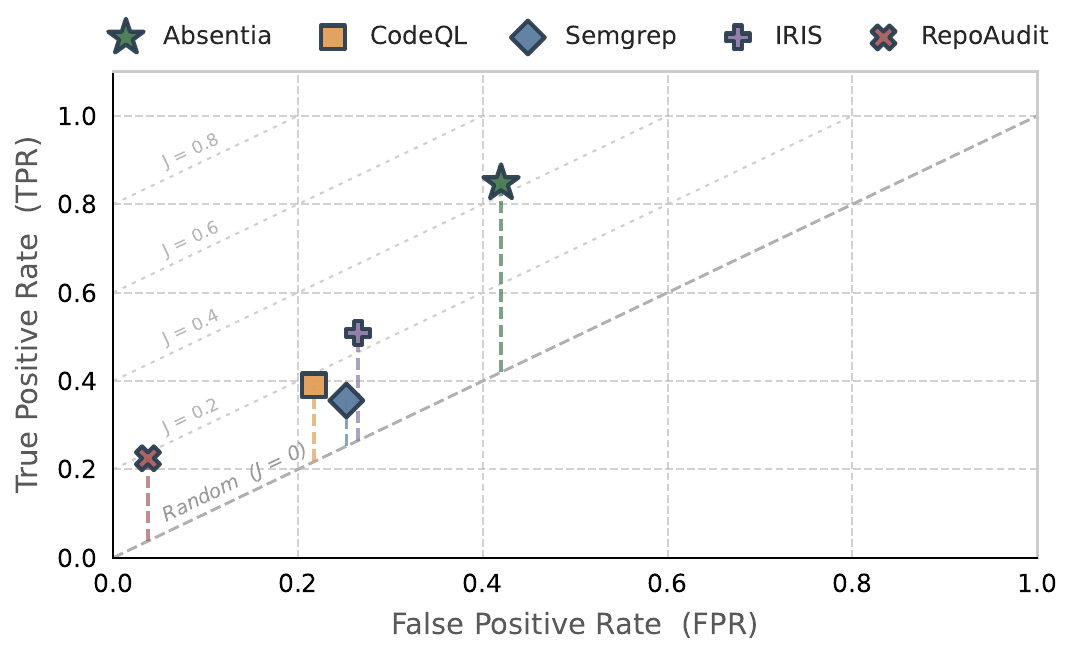}
        \caption{Python}
        \label{fig:owasp-injection-python}
    \end{subfigure}
    \hfill
    \begin{subfigure}[t]{0.49\linewidth}
        \centering
        \includegraphics[width=\linewidth]
            {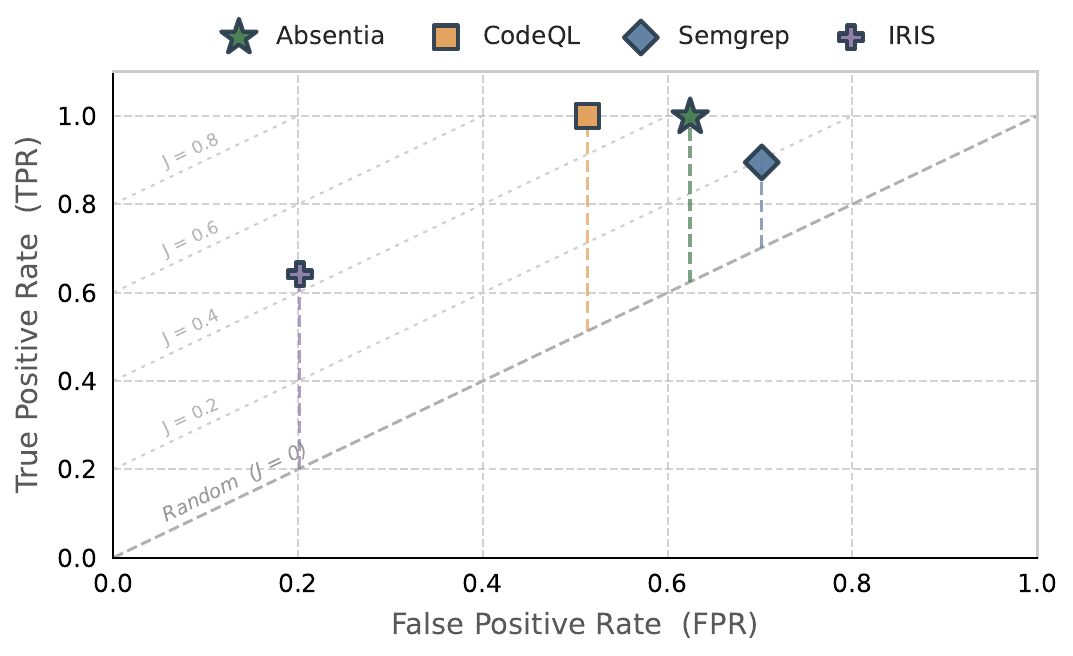}
        \caption{Java}
        \label{fig:owasp-injection-java}
    \end{subfigure}

    \caption{Injection detection on the OWASP Benchmarks, one marker per tool in ROC space, macro-averaged over the injection categories of each suite; a tool absent from a panel does not run on that language. The taint baselines are strongest on Java and fall back on Python, where \method leads them.}

    \label{fig:owasp-injection-scatter}
\end{figure}

Table~\ref{tab:ablation} leaves open whether \method reaches only that one class of defects, so we run the same configuration unchanged in the injection suites. Figure~\ref{fig:owasp-injection-scatter} places each tool in the ROC space, where Youden's~J is the vertical distance above the diagonal of chance. The baselines are strongest on Java and weaker on Python, and the ordering of the two panels differs: CodeQL leads on Java, 0.49 against \method's 0.37, and \method leads on Python, 0.43 against IRIS's 0.24. What the panels separate them on is how far each moves between the two: 0.31 for CodeQL, 0.20 for IRIS, 0.09 for Semgrep, and 0.06 for \method. Every query CodeQL and Semgrep run was written before they saw the code, and how well those queries fit is a property of the language they were written for. \method writes nothing down in advance: it derives what to match from the repository in front of it (\S\ref{sec:design-shared-fact-substrate}), so a language for which it does not have queries is not a language on which it is weaker, which is why \method ends up working in multiple languages.

\par
The markers also say what kind of error each tool makes, and the two model-driven baselines make the one they were built to make. IRIS adds a model to a taint analysis in order to discard the flows it should not have reported, and on Java it holds the lowest false-positive rate in the panel, 0.20 against CodeQL's 0.51 on the same queries it compiles into, while RepoAudit goes further still, to 0.04. What neither reaches is the vulnerable cases, at true-positive rates of 0.64 and 0.23.
\par
\method sits on the other side of both panels, and its error is not flattering either. It reaches a true-positive rate of 0.85 in Python and 1.00 in Java at false-positive rates of 0.42 and 0.62. The two positions are not equivalent for whoever runs the tool. A false positive costs a reviewer the time to read the case and dismiss it, while a vulnerability nothing reports stays in the application with no later stage to recover it, which makes it the costlier of the two errors to end up on. Appendix~\ref{sec:appendix-injection-model} substitutes Claude Sonnet-5 for MiniMax-2.5 on the Python suite, where the false-positive rate falls from 0.42 to 0.14 and the true-positive rate rises from 0.85 to 0.95 at 6.2 times the bill. Most of the observed FPR is therefore the model rather than the method, and MiniMax-2.5 remains the configuration every other measurement here is taken on.

\FloatBarrier
\subsection{Cost and Continuous Analysis (RQ4)}
\label{sec:eval:rq4}

\begin{figure}[htbp]
    \centering
    \includegraphics[width=0.76\linewidth]{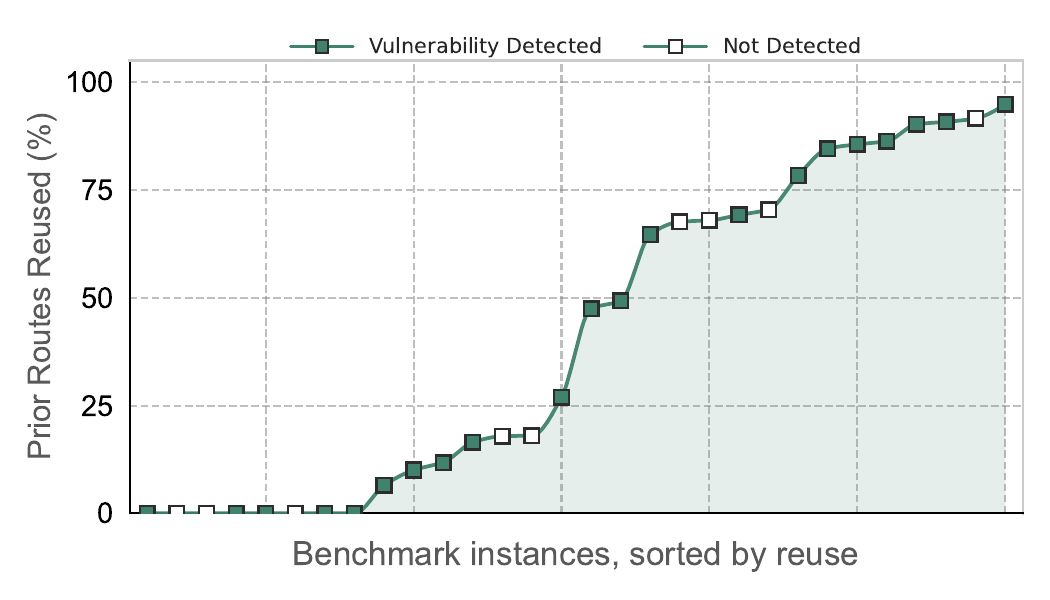}
    \captionsetup{width=.82\linewidth}
    \caption{Share of routes whose verdicts remain unchanged when a repository is re-analyzed at a later commit. Each marker represents one instance; filled markers indicate that the vulnerability is still reported.}
    \label{fig:reuse}
\end{figure}

At the MiniMax-2.5 rates of \S\ref{sec:implementation}, a run of \method costs a median of \$44 and between \$2 and \$316 across the benchmark, of which the per-route agents, at about \$0.20 a route, are 95\% of a median bill.
\par
Figure~\ref{fig:reuse} runs the re-analysis of \S\ref{sec:design-incremental} over all 30 instances, taking the fixed commit as the earlier analysis and the vulnerable one as the commit re-analyzed, which is the order that lets the disclosed defect be scored on the second of the two. A median commit reuses 37\% of the routes the previous run analyzed, and the curve is a split rather than a central tendency: 9 instances reuse 70\% or more, where nearly the whole bill is avoided, and 8 reuse nothing, since a change in a file most of the route table is built on invalidates most of the precomputed verdicts.
\par
A saving is only worth taking if the verdicts carried over are still the right ones, which is what scoring the vulnerable commit second establishes. Recall of the vulnerability is observed in 21 of the 30 instances, at 95\% reuse and at 0\% alike, which is above the 19 of Table~\ref{tab:authz-recall} rather than below it. The two sets differ by 4 instances gained and 2 lost, within the run-to-run variation Table~\ref{tab:reliability} measures, so the saving carries no detection penalty we can observe, which is what makes \method affordable to run over a sequence of commits rather than once.

\section{Discussion and Limitations}
\label{sec:discussion-limitations}

\myparagraph{Verification effort.} \method leaves on the order of 77 findings per benchmark instance, of which the verifier confirms about half (\S\ref{sec:eval:rq2}), against the 213 that Semgrep averages on the same repositories, none of which is the disclosed vulnerability. Fewer and denser as they are, a person still reads all of them, about two for every defect they confirm, and in no particular order: \method assigns its highest confidence to most of what it emits, which better calibrated could tell a reviewer where to start.

\myparagraph{Extraction coverage.} Although \method analyzes every route in its graph, that graph is the surface the extraction stage recovered and may not be the application's true one, so a route the patterns never match is absent from every stage after it.

\myparagraph{Nondeterminism.} While \method narrows the variation between its runs, judging every route at temperature zero and unioning two extraction passes so that a route one pass misses can still be recovered by the other (\S\ref{sec:implementation}), no such measure makes a language model deterministic, and \method will not reach the reproducibility of a static analyzer, which returns the same findings on every run over a commit. We did see, nonetheless, that on repeated trials, detection is mostly a property of the instance rather than of the draw, so nondeterminism does not appear to be fatal here.

\myparagraph{Exposure to the advisories.}\tinyquad \method runs on MiniMax-2.5, from the MiniMax M2 family. That family documents a knowledge cutoff of June 2025, though not for M2.5 itself. Every advisory in \bench was disclosed later, the earliest on 3 July 2025. We read \bench as outside the model's training data on that basis, without being able to establish it.

\myparagraph{Cost.} Reading every route is what makes coverage systematic, and it is also what \method spends its budget on, a median of \$44 per repository and up to \$316 on the largest (\S\ref{sec:eval:rq4}). The analyzers it is compared against carry no per-run cost at all, so this is not a margin to be optimized against them but the price of a capability they do not offer at any setting. Where it can be reduced is on re-analysis rather than on the first run, since a commit leaves the verdicts of the routes it cannot affect intact, and a median commit in our benchmark reuses 37\% of them.

\myparagraph{Scope of the evidence.} Our instances span nine frameworks and three languages, but we acknowledge that the space of web applications is far larger than thirty of them can cover, so the recall of \S\ref{sec:eval:rq1} may differ once \method is deployed in the wild. The benchmark is also silent on one stage of \method: no advisory in it requires two routes to be chained. Each localizes to a single handler, even where several routes reach it, so that recall owes nothing to the stage that composes findings across routes. That stage is not idle nonetheless. It composed 42 pairs over the thirty runs, 13 of which fell inside the sample \S\ref{sec:eval:rq2} verifies, and about half of those 13 were confirmed, the same share the verifier confirmed among the single-route findings of the same runs. We report composition as a feature of the pipeline rather than an adjudicated capability.

\section{Related Work}
\label{sec:related-work}

\myparagraph{Detecting vulnerabilities in web applications.}
Program analysis for discovering web vulnerabilities has been extensively studied, through taint tracking~\cite{sql-injection-livshits} and the query-based analyzers that run over a repository in continuous scanning~\cite{codeql,semgrep}. These are effective where a bug can be defined once for a language, namely as a path carrying untrusted input into a dangerous operation, but they rely on expert-written source and sink specifications and on manual review of the alerts they raise. No such portable definition exists for broken access control, since what a route may expose is particular to one application. Detectors for it therefore take a notion of the intended policy from elsewhere: a specification the developer supplies~\cite{static-analysis-usenix}, consistency with the guards of comparable operations (RoleCast~\cite{rolecast-sast-bac}, MACE~\cite{mace-sast-bac}), or a relation fixed in advance~\cite{bolaray-sast-bac}, as a pattern catalog (SPACE~\cite{space-sast-bac}) or as ownership read off the database schema (MOCGuard~\cite{mocguard2025}). Where the code is not read at all, authorization is instead tested by running the application and replaying one principal's requests under another's session~\cite{owasp-wstg-42,zap,bacscan,bacfuzz}, at the cost of a deployment, an account for every role, and a way to drive the application into the state where each check runs.

\myparagraph{Language models.}
Identifying a vulnerability was for some time beyond what a language model could do reliably, its verdicts being non-deterministic and sensitive to superficial edits such as a renamed function~\cite{secholmes,primevul}. Frontier models are considerably stronger at it, the most recent reporting thousands of zero-days in deployed software~\cite{mythos}. That earlier gap motivated a line of work pairing a model with a static analysis, so that each covers what the other does poorly: for example, by having a model write the taint specifications a query needs (IRIS~\cite{iris}), by having it filter the findings a reviewer would otherwise triage (LLift~\cite{llift-sast-fp-reduction}), or by abstracting a repository into a code property graph so that what the model reads fits its context (LLMxCPG~\cite{llmxcpg-sast}).

\myparagraph{Agents.}
Given tools and left to choose what to read, a model need not be paired with an analysis at all, and agents of this kind audit a repository on demand (RepoAudit~\cite{repoaudit}, Codex Security~\cite{codex-security}). Exploring code and reasoning about what it means, rather than matching a pattern on it, is what allows an agent to reach a bug whose definition was never written down, and access control has begun to be approached this way~\cite{neo}. Closest to \method is Co-RedTeam, whose discovery stage receives only the target codebase and has agents read it for candidates with supporting evidence~\cite{co-red-team}. However, it reports no implementation we can obtain or run. A parallel line establishes a vulnerability by exploiting it from the outside, an agent conducting a penetration test against a running deployment and planning its way along an attack chain~\cite{pentestgpt,pentestagent,pentestgpt-v2}. Approaching an application in this way requires one to be running. \method reads its source instead, in the setting of \S\ref{sec:scope}, where the engineers who own the code audit it before it is deployed.

\section{Conclusion}
\label{sec:conclusion}

Checking authorization in source code is a reading task, and we show that this reading can be automated: the property a route is meant to guarantee is recoverable from the application's own code, and that code can then be held to it, so a reviewer checks a claim instead of performing the analysis. On \bench, the benchmark of 30 disclosed advisories we release, \method reports 19, where the static analyzers we run report none and an unstructured agent on the same model reports 3. Recovering a property is also more general than having one written in advance, so a single unchanged configuration reaches both injection, which is defined once per language, and broken access control, which is defined once per application.

\clearpage
\section*{Ethical Considerations}
\label{sec:ethical-considerations}

\paragraph{Subjects and data.} Our evaluation targets open-source applications, and every instance of the benchmark of \S\ref{sec:benchmark} corresponds to a vulnerability that its maintainers had publicly disclosed and fixed before this study began, while the injection suites of \S\ref{sec:setup} are synthetic by construction. \method analyzes application source code and does not execute the applications it examines (\S\ref{sec:scope}), so no system belonging to a third party was accessed, and no data belonging to the users of any application was collected.

\paragraph{Findings beyond the disclosed advisories.} \method reported findings other than the thirty advisories under test, and \S\ref{sec:eval:rq2} verified a sample of them against the code. We have not triaged the remainder and did not forward them to maintainers. Automated findings at this volume impose a real cost on the volunteers who would receive them, and a set of unvetted model-generated reports is not a responsible disclosure. We intend to triage the highest-confidence findings under expert review and to disclose those that survive through each project's stated security channel. Until that review is done we treat them as candidates rather than as vulnerabilities, and no claim in this paper depends on them.

\paragraph{Human evaluation.} The manual validations reported in \S\ref{sec:benchmark} and \S\ref{sec:eval:rq2} were performed by one of the authors, a graduate researcher in security, and no external participants were recruited or compensated. To avoid bias, that author carried out each task under the same blinding as an independent reviewer would have had, without access to the verdict under review or to which tool or configuration produced the finding.

\paragraph{Potential for misuse.} We acknowledge that a method of this kind, and the findings it produces, could be repurposed to locate vulnerabilities for malicious ends. \method is intended for research and for the review of code that the user owns or is authorized to audit, and we ask that it be used accordingly.

\section*{Open Science}
\label{sec:open-science}

Code and data will be available on:
\url{https://github.com/avduarte333/Absentia}.

\section*{Acknowledgments}
\label{sec:acknowledgments}

This work was conducted as part of the Amazon Nova AI Challenge, which provided funding and AWS compute credits to participating university teams. The study design, model selection, analysis, and the decision to publish were made independently by the authors. We also acknowledge support from national funds through Fundação para a Ciência e a Tecnologia, I.P. (FCT), under projects UID/50021/2025 and UID/PRR/50021/2025. This work is also co-financed by FCT through the Carnegie Mellon Portugal Program under the fellowship PRT/BD/155049/2024. Li is partly supported by the CMU CyLab seed grant.

\clearpage
\bibliographystyle{plainurl}
\bibliography{references}

\clearpage
\appendix
\section{The Benchmark Instances}
\label{sec:appendix}
Table~\ref{tab:instances} lists the 30 instances one by one, where \S\ref{sec:benchmark} characterizes them in aggregate.
\begin{table}[htbp]
  \centering
  \caption{The 30 instances of the authorization benchmark grouped by implementation language, where \emph{released} is the month the advisory was published, \emph{CWE} is the first weakness class the advisory assigns, \emph{routes} estimates the request entry points each repository declares at the vulnerable commit, independently of \method's extraction, and advisory identifiers omit the constant \texttt{GHSA-} prefix. The release dates span July 2025 to June 2026.}
  \label{tab:instances}
  \small
  \setlength{\tabcolsep}{8pt}
  \begin{tabular*}{\linewidth}{@{\extracolsep{\fill}}llllrr@{}}
    \toprule
    \textbf{Repository} & Advisory & Date & Lang. & CWE & Routes \\
    \midrule
    FUXA & 8GHR-W65F-J3QR & Jun'26 & JS & 862 & 59 \\
    parse-server & 5F92-JRQ3-28RC & Mar'26 & JS & 862 & 89 \\
    saltcorn & 9237-RG5P-RHFW & Apr'26 & JS & 863 & 422 \\
    \addlinespace
    aegra & M98R-6667-4WQ7 & May'26 & Py & 639 & 46 \\
    askbot-devel & R2JV-FWFR-4J8C & Jan'26 & Py & 639 & 156 \\
    indico & Q28V-664F-Q6WJ & Jul'25 & Py & 639 & 999 \\
    khoj & 6WHJ-7QMG-86QJ & Feb'26 & Py & 639 & 98 \\
    langflow & RF6X-R45M-XV3W & Mar'26 & Py & 639 & 162 \\
    langflow & 8C4J-F57C-35CF & Mar'26 & Py & 639 & 126 \\
    langflow & 7GRX-3XCX-2XV5 & Mar'26 & Py & 862 & 206 \\
    mlflow & 46R5-X6JQ-V8G6 & Apr'26 & Py & 862 & 193 \\
    open-webui & 26G9-27VM-X3Q8 & May'26 & Py & 639 & 462 \\
    plane & QW87-V5W3-6VXX & May'26 & Py & 639 & 682 \\
    pretix & PMJJ-H5JM-VXH4 & Dec'25 & Py & 639 & 878 \\
    recipes & V8X3-W674-55P5 & Apr'26 & Py & 639 & 368 \\
    wagtail & 4QVV-G3VR-M348 & Feb'26 & Py & 862 & 217 \\
    weblate & WPPC-7CQ7-CGFV & Feb'26 & Py & 862 & 407 \\
    weblate & MPF5-3VPH-Q75R & Apr'26 & Py & 862 & 410 \\
    weblate & 3PMH-24WP-XPF4 & Dec'25 & Py & 284 & 380 \\
    wger & G8GC-6C4H-JG86 & Feb'26 & Py & 639 & 420 \\
    wger & 42CR-W2GR-M54Q & Feb'26 & Py & 639 & 420 \\
    \addlinespace
    actual & QMJJ-P7M9-WJRV & Feb'26 & TS & 862 & 51 \\
    Flowise & 3PRP-9GF7-4RXX & Apr'26 & TS & 284 & 234 \\
    git-proxy & 39P2-8HQ9-FWJ6 & Jul'25 & TS & 285 & 30 \\
    n8n & GQ57-V332-7666 & Jul'25 & TS & 862 & 240 \\
    outline & 23JJ-RP48-W7Q7 & Apr'26 & TS & 639 & 245 \\
    payload-puck & 65W6-PF7X-5G85 & Apr'26 & TS & 862 & 17 \\
    rallly & 4J32-25F9-QGFM & Nov'25 & TS & 285 & 95 \\
    strapi & 495J-H493-42Q2 & Oct'25 & TS & 639 & 224 \\
    studiocms & 9V82-XRM4-MP52 & Mar'26 & TS & 639 & 112 \\
    \bottomrule
  \end{tabular*}
\end{table}

\clearpage

\section{Injection Under Sonnet-5}
\label{sec:appendix-injection-model}

The false-positive rate \method records on the injection suites, 0.42 in Python and 0.62 in Java (\S\ref{sec:eval:rq3}), could be a consequence of inspecting every route rather than of the model reading them. The authorization benchmark cannot tell the two apart, because a rate needs safe cases and real applications do not supply them (\S\ref{sec:setup}). We therefore repeat the Python run with Claude Sonnet-5 at every model-driven stage of \method, in place of the MiniMax-2.5 every other measurement in this paper is taken on.

\begin{table}[htbp]
  \centering
  \captionsetup{width=.90\linewidth}
  \caption{True-positive and false-positive rate per injection category on the OWASP Benchmark Python under MiniMax-2.5 and under Claude Sonnet-5.}
  \label{tab:injection-model}
  \small
  \setlength{\tabcolsep}{8pt}
  \begin{tabular*}{.90\linewidth}{@{\extracolsep{\fill}}l cc cc r@{}}
    \toprule
    & \multicolumn{2}{c}{MiniMax-2.5} & \multicolumn{2}{c}{Sonnet-5} & \\
    \cmidrule(lr){2-3}\cmidrule(lr){4-5}
    \textbf{Category} & TPR & FPR & TPR & FPR & $\Delta$FPR \\
    \midrule
    SQL injection          & 0.60 & 0.00 & 1.00 & 0.00 & $+0.00$ \\
    Command injection      & 0.92 & 0.43 & 1.00 & 0.29 & $-0.14$ \\
    LDAP injection         & 0.69 & 0.54 & 1.00 & 0.46 & $-0.08$ \\
    XPath injection        & 1.00 & 0.60 & 1.00 & 0.10 & $-0.50$ \\
    Cross-site scripting   & 0.97 & 0.41 & 1.00 & 0.09 & $-0.33$ \\
    Path traversal         & 0.98 & 0.46 & 1.00 & 0.14 & $-0.32$ \\
    Code injection         & 0.45 & 0.45 & 0.50 & 0.09 & $-0.36$ \\
    Unsafe deserialization & 0.94 & 0.22 & 1.00 & 0.03 & $-0.19$ \\
    Unchecked redirect     & 0.92 & 0.33 & 1.00 & 0.10 & $-0.24$ \\
    XML external entity    & 1.00 & 0.75 & 1.00 & 0.10 & $-0.65$ \\
    \textbf{Macro-average} & \textbf{0.85} & \textbf{0.42} & \textbf{0.95} & \textbf{0.14} & $-0.28$ \\
    \bottomrule
  \end{tabular*}
\end{table}

Each run enumerated the same 2{,}463 route request surface, and both are scored by the same scorer over the same 677 injection cases, so what separates the two pairs of columns is the model. Sonnet-5 emits 798 findings across the suite against MiniMax-2.5's 1{,}258, so the lower false-positive rate is a better discriminated set of findings.

\end{document}